\documentclass[3p,times,sort&compress]{elsarticle}

\usepackage{ecrc}

\usepackage{epstopdf}
\usepackage{afterpage}
\usepackage{hyperref}

\volume{00}

\firstpage{1}

\journalname{Nuclear Physics A}

\runauth{Anton Andronic}

\jid{nupha}

\jnltitlelogo{Nuclear Physics A}

\usepackage{graphicx}
\usepackage{amsmath,amssymb}

\newcommand{\ud}{\mathrm{d}}

\begin{document}

\begin{frontmatter}

\title{Experimental results and phenomenology of quarkonium production
in relativistic nuclear collisions}

\author{Anton Andronic}
\address{Reasearch Division and EMMI, GSI Helmholtzzentrum f\"ur Schwerionenforschung, D-64291 Darmstadt, Germany}

\begin{abstract}
An overview of recent measurements of quarkonium production in nucleus-nucleus 
collisions and their understanding in theoretical models is given.
\end{abstract}


\end{frontmatter}


\section{Introduction}
\label{intro}

Among the various suggested probes of deconfinement, charmonium ($c\bar{c}$) 
states play a distinctive role. The J/$\psi$ meson is the first hadron for 
which a clear mechanism of suppression in deconfined matter (``quark-gluon 
plasma", QGP) was proposed based on the color analogue of Debye 
screening \cite{Matsui:1986dk}. 
Current terminology includes 
``dissociation" and ``melting", which are somewhat harder terms  
compared to the initial proposal of screening as ``hindered binding" of $c$ and $\bar{c}$ 
quarks.
Further refinements, including all quarkonium species, $c\bar{c}$ and
$b\bar{b}$, led to the picture of ``sequential suppression" 
\cite{Karsch:1990wi,Digal:2001ue,Karsch:2005nk}: a hierarchy of quarkonium 
dissociation dependent on the binding energy (size) of the quarkonium state, 
which could give information on the temperature of the medium, given that 
the Debye length in deconfined matter has a pronounced temperature 
dependence \cite{Digal:2001ue}.
It was pointed out early-on that the Debye screening phenomenon is a 
low-$p_{\rm T}$ effect \cite{Blaizot:1988ec,Karsch:1990wi}, an issue 
highlighted in current research \cite{Kopeliovich:2014uha}.
Lattice QCD calculations can give information on the screening 
\cite{Asakawa:2003re,Datta:2003ww,Mocsy:2007yj,Mocsy:2007jz},
a subject of intense current research
\cite{Morita:2010pd,Ding:2012sp,Hayata:2012rw,Lee:2013dca,Rothkopf:2013kya,Borsanyi:2014vka,Aarts:2014cda} (see a review in \cite{Mocsy:2013syh}).
The theoretical challenge \cite{Laine:2011xr} is illustrated by the spread of 
results obtained with various approaches \cite{Adare:2014hje}.
A review of charmonium data at the SPS and RHIC and its interpretation in the 
screening scenario is available in ref. \cite{Kluberg:2009wc}.

Novel quarkonium production mechanisms were proposed in the year 2000.
In the statistical hadronization model \cite{BraunMunzinger:2000px}, the 
charm (bottom) quarks and antiquarks, produced in initial hard collisions, 
thermalize in QGP and are ``distributed'' into hadrons at chemical freeze-out. 
It is assumed that no quarkonium state is produced in the deconfined 
state (full suppression). Quarkonia are produced, together with all other 
hadrons, at chemical freeze-out (hadronization) \cite{BraunMunzinger:2000px,Andronic:2006ky}. A variant with partial suppression of initial charmonium 
production has been also proposed \cite{Grandchamp:2001pf}.
An important aspect in this scenario is the canonical suppression of open charm
or bottom hadrons \cite{BraunMunzinger:2000ep,Gorenstein:2001xp}, which 
determines the centrality dependence of production yields in this model. 
The overall magnitude is determined by the input charm (bottom) production 
cross section \cite{Andronic:2007bi}.
See a review in \cite{BraunMunzinger:2009ih} and more recent 
predictions of this model in \cite{Andronic:2011yq} and of a similar 
approach in \cite{Ferreiro:2012rq}.

Kinetic (re)combination of heavy quarks and antiquarks in QGP \cite{Thews:2000rj}
is an alternative quarkonium production mechanism.
In this transport model (see ref. \cite{Liu:2009nb,Zhao:2011cv,Emerick:2011xu,Zhou:2014kka} 
for recent results) there is continuous dissociation and (re)generation of 
quarkonium over the entire lifetime of the deconfined stage. 
A hydrodynamical-like expansion of the fireball of deconfined matter, 
constrained by data, is part of the model, alongside an implementation of 
the screening mechanism with inputs from lattice QCD.  
Other important ingredients are parton-level cross sections and, as in the 
case of the statistical hadronization model, the production cross section of 
initial $c\bar{c}$ ($b\bar{b}$) pairs and quarkonium states.

A wealth of data on charmonium and bottomonium production in nucleus-nucleus 
collisions has become recently available at RHIC 
\cite{Adare:2011yf,Adamczyk:2012pw,Adamczyk:2012ey,Adamczyk:2013tvk,Adamczyk:2013poh,Aidala:2014bqx,Adare:2014hje}
and at the LHC \cite{Abelev:2012rv,Chatrchyan:2012np,Chatrchyan:2012lxa,CMS-PAS-HIN-12-014,ALICE:2013xna,CMS:2013dla,Abelev:2013ila,Abelev:2014nua}, 
significantly extending our understanding of quarkonium production 
in deconfined matter.
The new data in proton (deuteron) collisions with nuclei, both at RHIC 
\cite{Adare:2013ezl,Adare:2014hje}
and at the LHC 
\cite{Abelev:2013yxa,Aaij:2013zxa,Chatrchyan:2013nza,Abelev:2014zpa,Aaij:2014mza,Arnaldi:2014kta}, are also significant. 
Initially intended to quantify the so-called ``cold nuclear effects",
effects of nuclear collisions not associated with hot (deconfined) matter, 
namely shadowing/saturation at collider energies, these data have revealed 
interesting aspects of quarkonium production.

\section{Charmonium}
\label{charm}

The measurement of J/$\psi$ production in Pb--Pb collisions at the LHC was 
expected to be decisive in clarifying the question of suppression via 
the Debye screening mechanism and answering what (re)generation 
scenarios are viable production mechanisms.
The data measured at high $p_{\mathrm T}$ \cite{Chatrchyan:2012np} showed a 
pronounced suppression of J/$\psi$ in Pb--Pb compared to pp collisions
and of the same magnitude as that of open-charm hadrons. 
This may indicate that the high-$p_{\mathrm T}$ charm quarks that form either 
$D$ or J/$\psi$ mesons had the same dynamics, determined by the energy loss 
process in deconfined matter.

\begin{figure}[hbtp]
\begin{tabular}{lr} \begin{minipage}{.49\textwidth}
\hspace{-0.3cm}\includegraphics[width=1.02\textwidth]{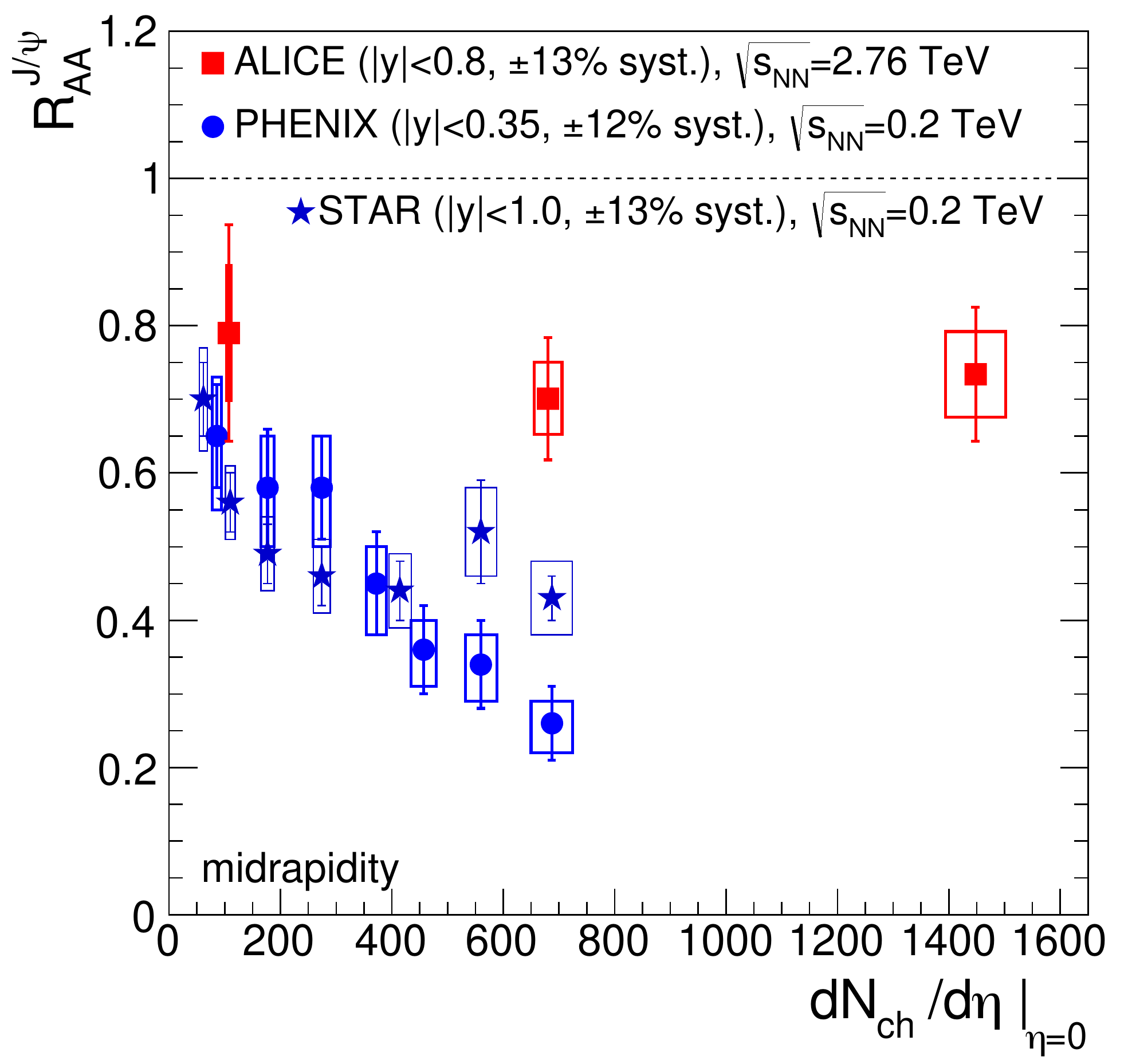}
\end{minipage} & \begin{minipage}{.49\textwidth}
\hspace{-0.3cm}\includegraphics[width=1.02\textwidth]{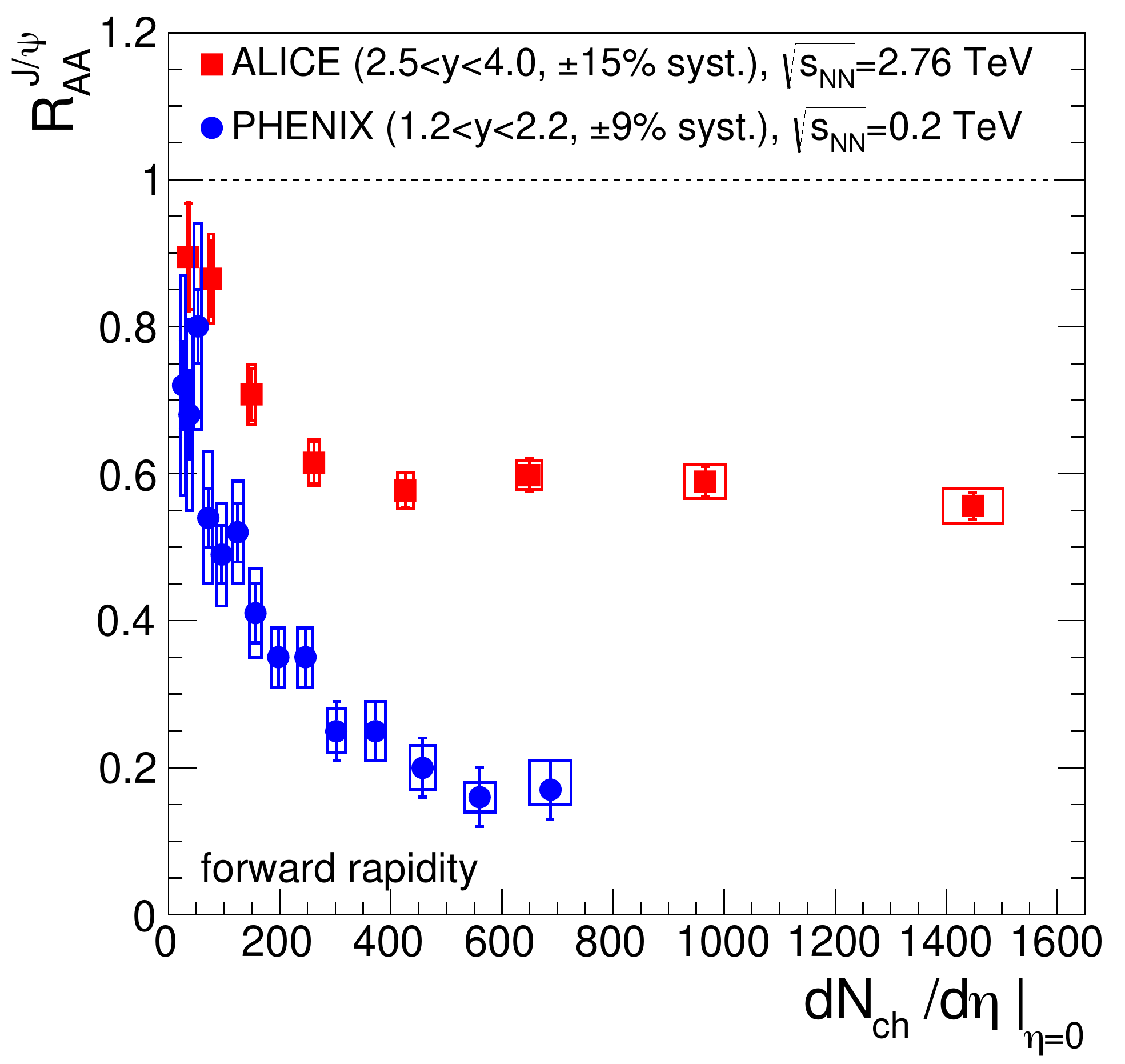}
\end{minipage}\end{tabular}
\caption{The dependence of the nuclear modification factor $R_{\mathrm{AA}}$ for
inclusive J/$\psi$ production on the charged-particle pseudorapidity density 
$\ud N_{\rm ch}/\ud\eta$ (at $\eta$=0) at midrapidity (left panel) and at 
forward rapidity (right panel).
The data are integrated over $p_{\mathrm T}$  and are from the PHENIX 
\cite{Adare:2006ns,Adare:2011yf} and STAR \cite{Adamczyk:2013tvk}
collaborations at RHIC and the ALICE collaboration \cite{Abelev:2013ila} 
at the LHC.
Note the additional systematic uncertainties of the data quoted in 
the legend.
}
\label{fig:raa_jpsi1}
\end{figure}

The first LHC measurement of the overall (inclusive in $p_{\mathrm T}$) 
production \cite{Abelev:2012rv}, showed at forward rapidities values of the 
nuclear modification factor $R_{\mathrm{AA}}$  significantly larger than those 
measured at RHIC energies.
The full-statistics data of Run-1 confirmed this, see Fig.~\ref{fig:raa_jpsi1},
where the LHC data \cite{Abelev:2013ila} are compared to RHIC data 
\cite{Adare:2006ns,Adare:2011yf,Adamczyk:2013tvk}
at midrapidity and forward rapidity. The comparison is performed as a function
of the charged particle pseudorapidity density $\ud N_{\rm ch}/\ud\eta$, measured
around $\eta$=0, which is a measure of the energy density of the system.
In the Debye screening scenario, the larger energy density reached at the LHC
is expected to lead to a reduced production of charmonium. In addition, 
the larger parton shadowing/saturation expected at the LHC implies lower 
$R_{\mathrm{AA}}$ values compared to the RHIC energy.
The opposite is observed, demonstrating that novel production mechanisms, 
beyond (in addition to) the Debye screening effects are at work.
This is a generic expectation of both statistical hadronization and kinetic 
(re)combination models, which predicted \cite{Andronic:2007bi,Liu:2009nb,Zhao:2011cv} 
larger $R_{\mathrm{AA}}$ values at the LHC compared to RHIC.

\begin{figure}[hbtp]
\centering\includegraphics[width=.55\textwidth,height=.43\textwidth]{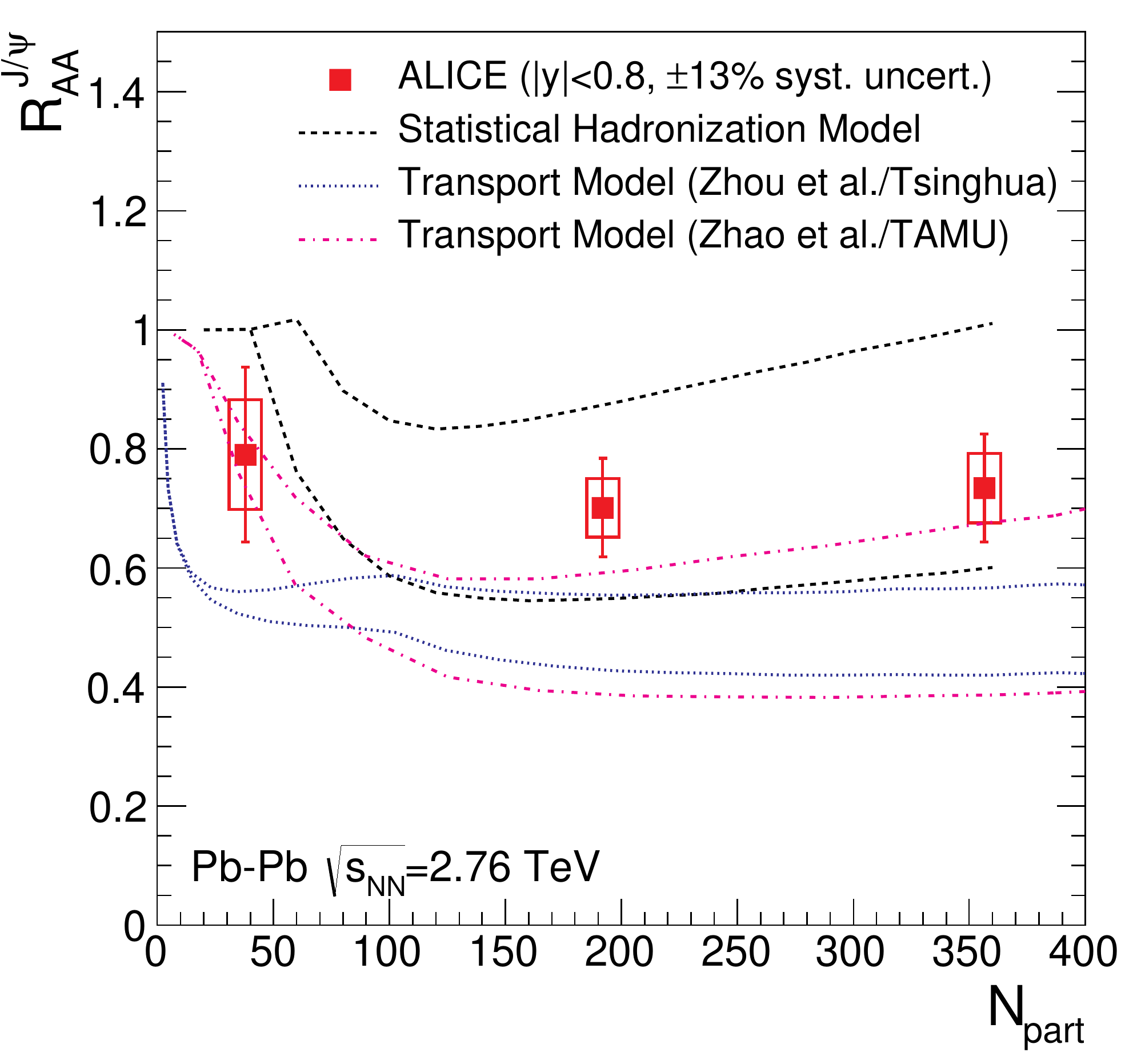}
\caption{Centrality dependence of the nuclear modification factor for inclusive 
J$/\psi$ production  at the LHC. The data at midrapidity \cite{Abelev:2013ila} 
are compared to model calculations: the statistical hadronization model 
\cite{Andronic:2011yq} and transport models of the TAMU \cite{Zhao:2011cv,Zhao:2012gc} 
and Tsinghua \cite{Liu:2009nb,Zhou:2014kka} groups. The bands are denoting
part of the uncertainty on the charm production cross section.
} 
\label{fig:raa_jpsi2} 
\end{figure}

The data are well described by the statistical hadronization model 
\cite{Andronic:2011yq} and by transport models \cite{Zhao:2011cv,Zhou:2014kka}, 
as illustrated in Fig.~\ref{fig:raa_jpsi2} for midrapidity data at the LHC.
This shows that J/$\psi$ production is a probe of deconfinement, confirming 
the initial idea \cite{Matsui:1986dk}, but may not be a simple  
``thermometer'' of the medium as initially hoped. 
Within the statistical model, the charmonium states become probes 
of the phase boundary between the deconfined and the hadronic phases. 
This extends the family of quarks employed for the determination of 
the hadronization temperature via the conjectured connection to the 
chemical freeze-out temperature extracted from fits of statistical model 
calculations to yields of hadrons with $u$, $d$, and $s$ quarks 
\cite{Andronic:2008gu}.

\begin{figure}[hbtp]
\begin{tabular}{lr} \begin{minipage}{.49\textwidth}
\centering\includegraphics[width=1.\textwidth]{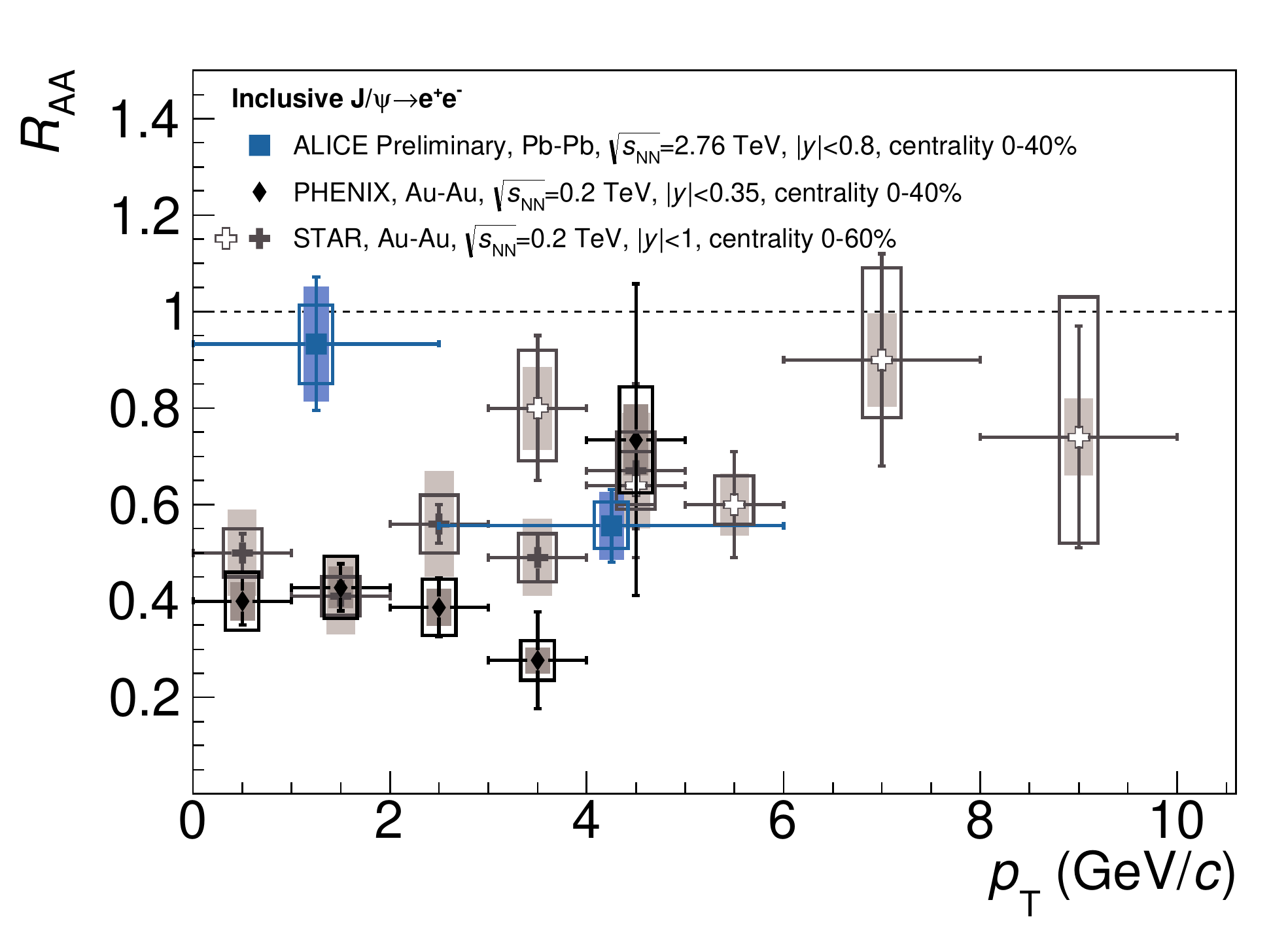}
\end{minipage} & \begin{minipage}{.49\textwidth}
\centering\includegraphics[width=1.\textwidth]{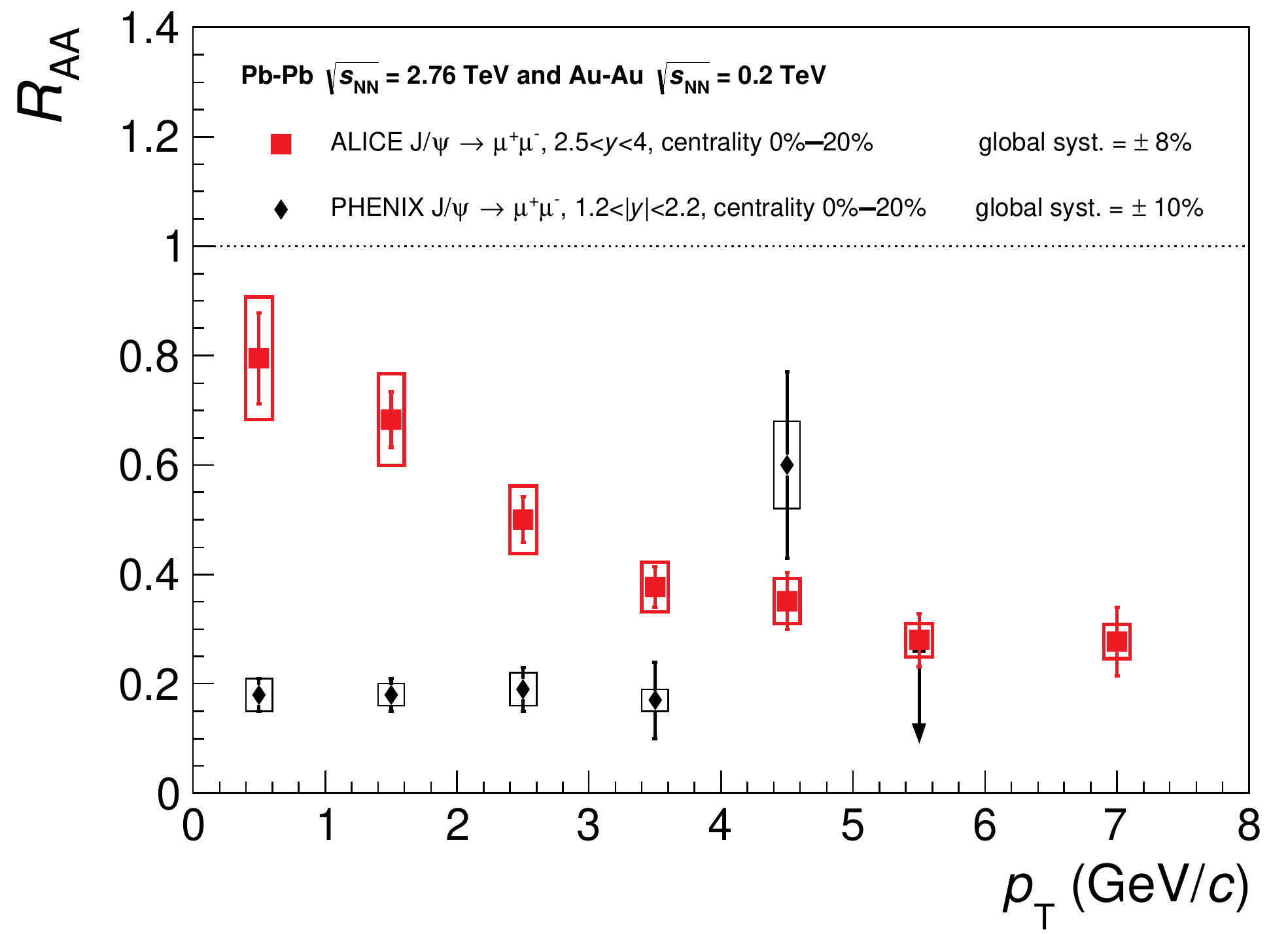}
\end{minipage}\end{tabular}
\caption{Transverse momentum dependence of the nuclear modification factor 
for J/$\psi$ production in central collisions measured at RHIC 
\cite{Adare:2006ns,Adare:2011yf,Adamczyk:2013tvk} and at the LHC \cite{Abelev:2013ila,Book:2014kta} 
at midrapidity (left panel) and at forward rapidity (right panel; 
plot from ref. \cite{Abelev:2013ila}).
} 
\label{fig:raa_jpsi3} 
\end{figure}

In transport models, about 60-80\% of the J/$\psi$ yield in central Pb--Pb 
collisions at $\sqrt{s_{_{\rm NN}}}=2.76$ TeV originates from (re)combination of 
$c$ and $\bar{c}$ quarks, the rest being primordial J/$\psi$ mesons that have 
survived in the deconfined medium.
In Au--Au collisions at $\sqrt{s_{_{\rm NN}}}=200$ GeV that fraction is about 
30-50\%.
It is worth pointing out that the input $c\bar{c}$ production cross section
in the statistical and transport models differs, at the LHC, by almost 
a factor of 2; further understanding of this difference and better constraints 
on $\sigma_{c\bar{c}}$ from data are needed in order to disentangle the
two competing models.
The study of J/$\psi$ yield relative to the inclusive $c\bar{c}$ yield was 
advocated \cite{Andronic:2006ky,Satz:2013ama}; to date, this is limited by the 
large uncertainties in the experimental determination of $\sigma_{c\bar{c}}$ 
in nucleus-nucleus collisions.

\begin{figure}[hbtp]
\begin{tabular}{cc} \begin{minipage}{.49\textwidth}
\hspace{-0.3cm}{\includegraphics[width=1.02\textwidth]{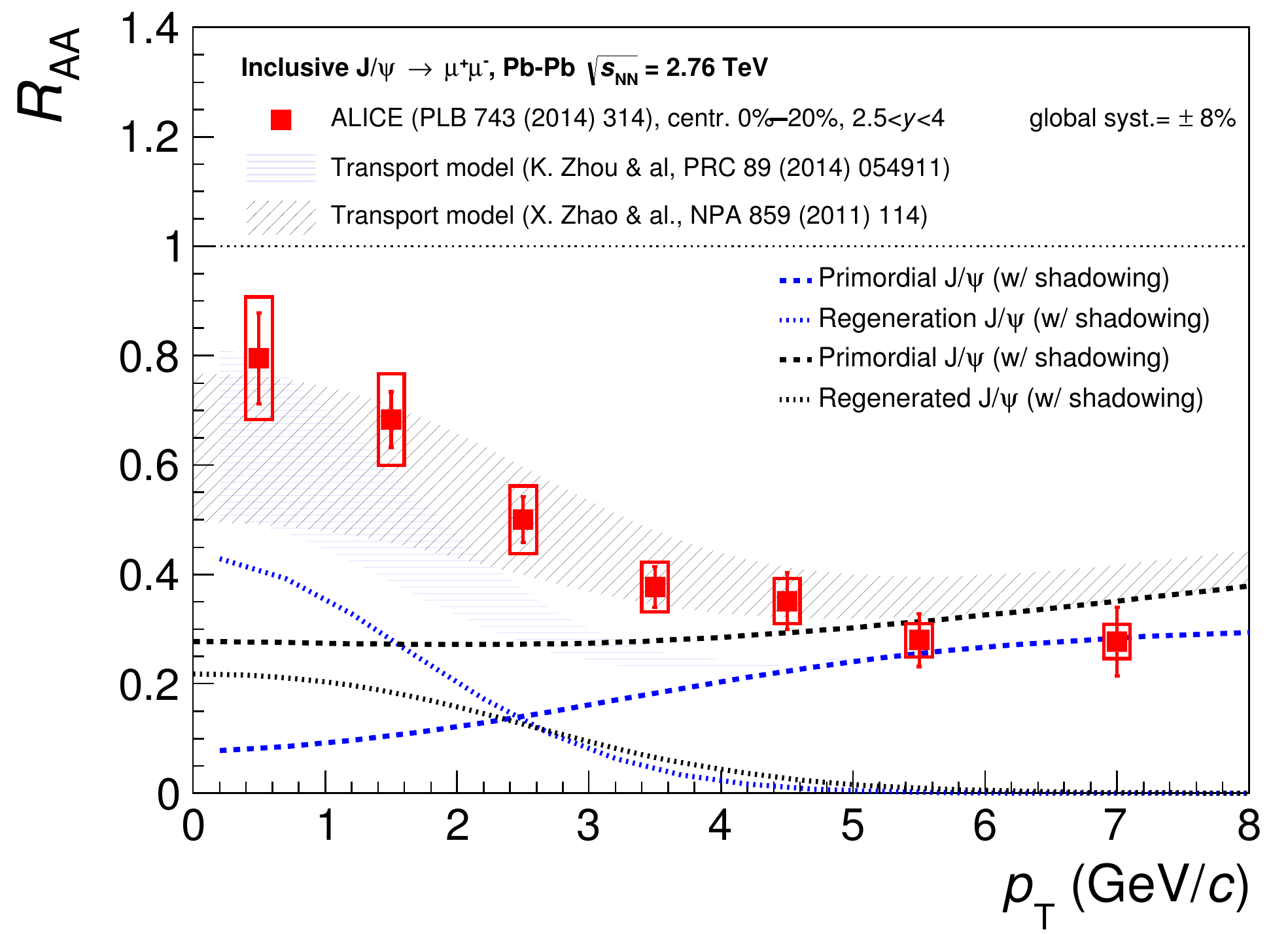}}
\end{minipage} &\begin{minipage}{.49\textwidth}
\hspace{-0.3cm}{\includegraphics[width=1.0\textwidth]{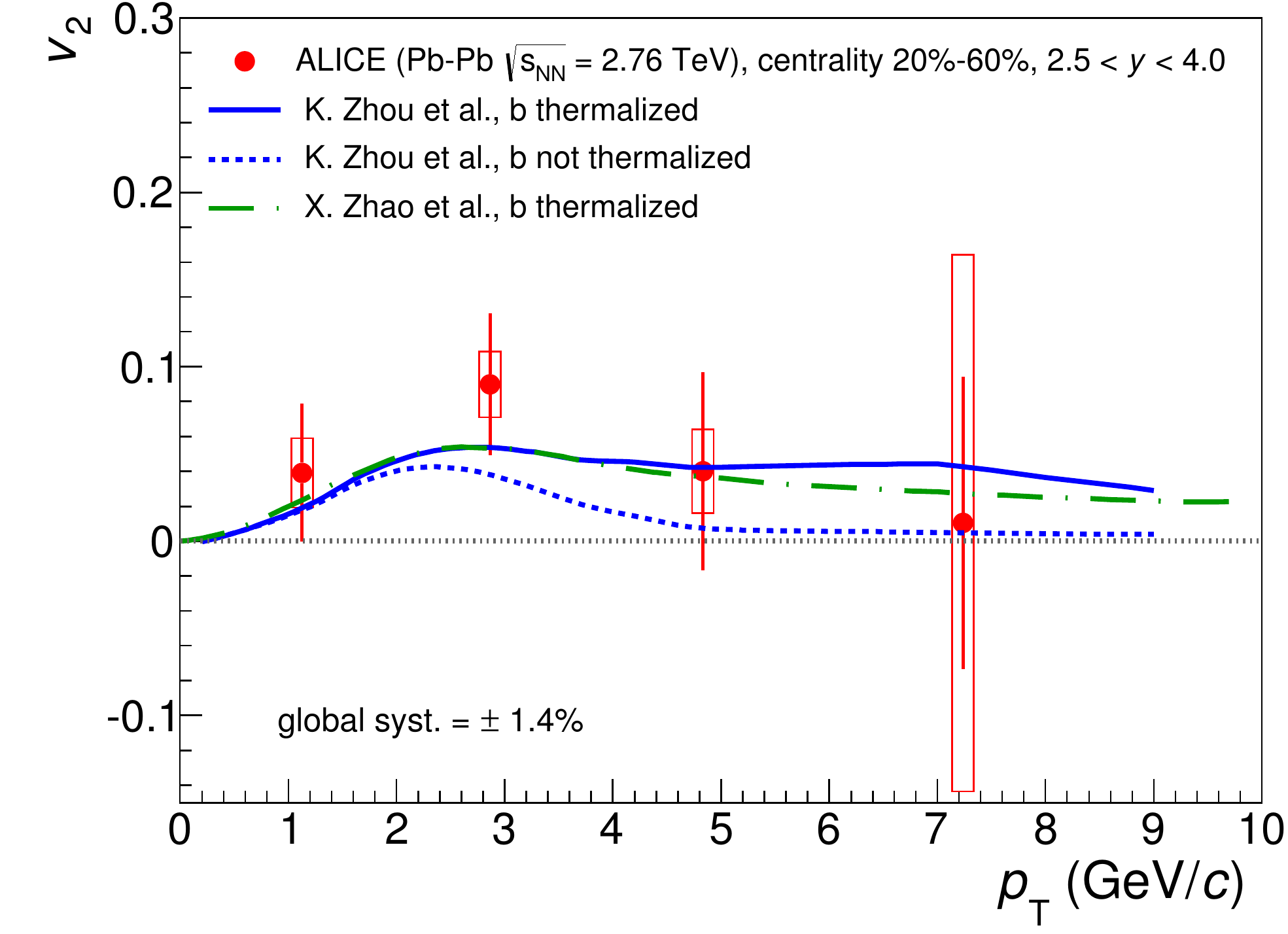}}
\end{minipage} \end{tabular}
\caption{Transverse momentum dependence of the nuclear modification factor 
(left panel \cite{Abelev:2013ila}) and elliptic flow coefficient $v_2$ 
(right panel \cite{ALICE:2013xna}) of J/$\psi$. 
The ALICE data at the LHC are at forward rapidity and are compared to transport 
model predictions of the TAMU (Zhao et al. \cite{Zhao:2011cv,Zhao:2012gc}) 
and Tsinghua (Zhou et al. \cite{Zhou:2014kka}) groups.
} 
\label{fig:raa_v2} 
\end{figure}

Another dramatic difference between the LHC and the RHIC data on J/$\psi$ 
production is observed in the transverse momentum dependence of the nuclear 
modification factor, shown in Fig.~\ref{fig:raa_jpsi3}.
This is arguably the most prominent difference seen between the LHC and RHIC 
data in all the multitude of observables available to date. 
Recalling that the Debye screening mechanism is expected to be effective 
at low $p_{\rm T}$ \cite{Blaizot:1988ec,Karsch:1990wi}, the features seen
in Fig.~\ref{fig:raa_jpsi3} give strong support to the interpretation of 
J/$\psi$ production at the LHC as dominated by generation at the hadronization 
or by (re)generation throughout the QGP lifetime, as the statistical 
hadronization model or the kinetic models, respectively, imply.
It was shown with transport models \cite{Zhao:2011cv,Zhou:2014kka} that, 
as expected, (re)generation is predominantly a low-$p_{\mathrm T}$ phenomenon, 
as illustrated by the comparison of LHC data to model predictions in 
Fig.~\ref{fig:raa_v2} (left panel).
This translates to average values of $p_{\rm T}$ very different at the LHC 
compared to lower energies \cite{Zhou:2014kka}.

J/$\psi$ data in Cu--Cu \cite{Adamczyk:2013tvk} and 
Cu--Au \cite{Aidala:2014bqx} 
collisions at RHIC exhibit a suppression close in magnitude to that observed 
in Au--Au collisions. 
At lower RHIC energies, measurements \cite{Adare:2012wf} show a J/$\psi$ 
suppression similar in magnitude to that measured at the top RHIC energy. 
This is in agreement with earlier measurements at the SPS and was implicitly
predicted within the statistical hadronization model \cite{Andronic:2006ky}; 
it is described in transport models \cite{Zhao:2010nk,Adare:2012wf}.

The measurement of J/$\psi$ elliptic flow at the LHC \cite{ALICE:2013xna,CMS:2013dla} 
brings another argument in favor of charmonium production from thermalized 
$c$ and $\bar{c}$ quarks. 
The transport model predictions \cite{Liu:2009gx,Zhao:2012gc} 
describe well the data, as shown in Fig.~\ref{fig:raa_v2} (right panel).
The recent measurement by the CMS collaboration \cite{CMS:2013dla} indicates 
that prompt J/$\psi$ mesons exhibit elliptic flow for $p_{\rm T}$ as large as 
10 GeV/$c$.  
We recall that the J/$\psi$ data at RHIC are compatible with a null flow 
signal \cite{Adamczyk:2012pw}.
A $v_2$ signal was measured for J/$\psi$  at the SPS \cite{Prino:2008zz} and 
was interpreted as a path-length dependence of the screening.

\begin{figure}[hbtp]
\begin{tabular}{cc} \begin{minipage}{.49\textwidth}
\hspace{0.1cm}{\includegraphics[width=.95\textwidth]{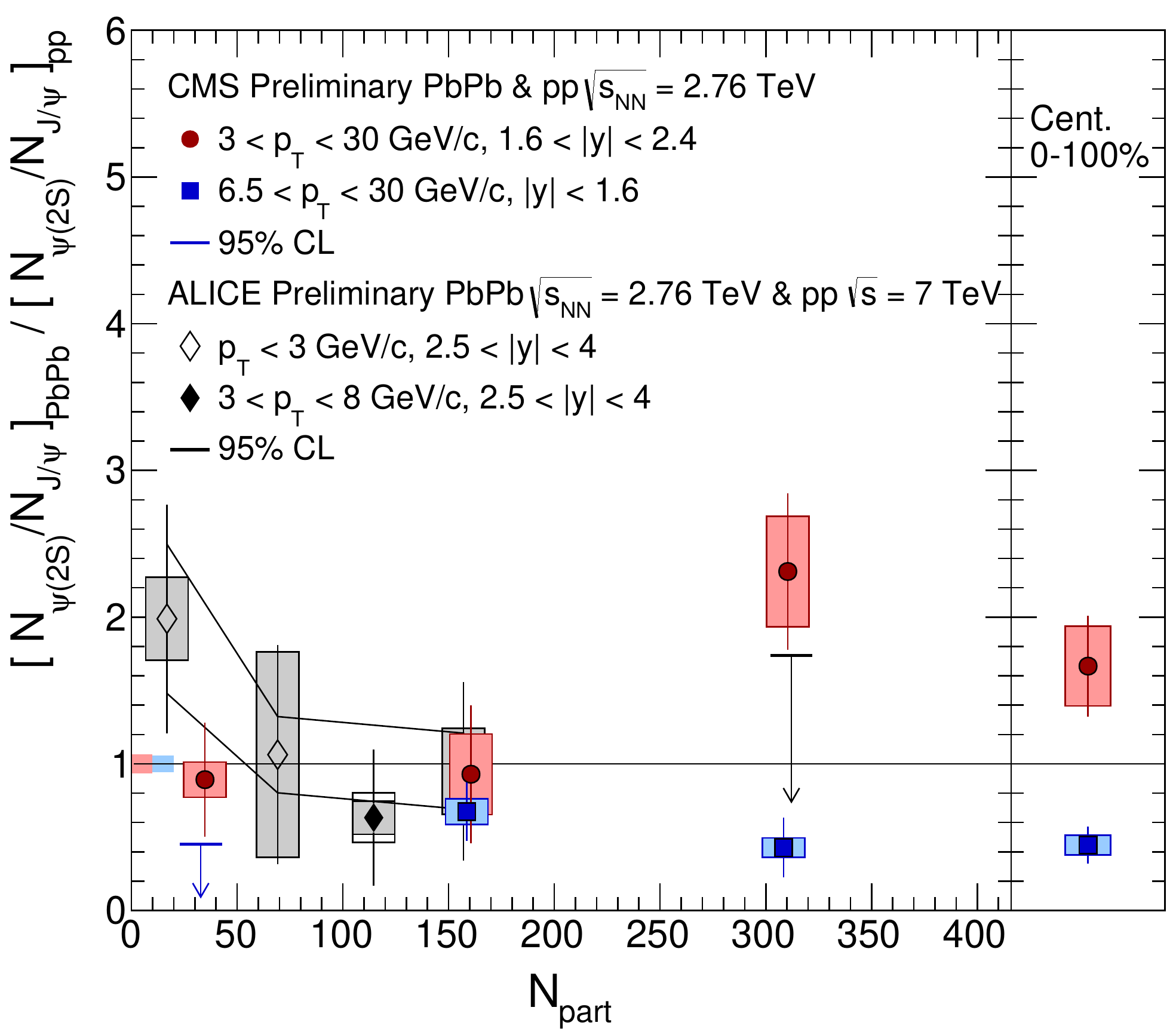}}
\end{minipage} &\begin{minipage}{.49\textwidth}
\hspace{-0.3cm}{\includegraphics[width=.95\textwidth,height=.82\textwidth]{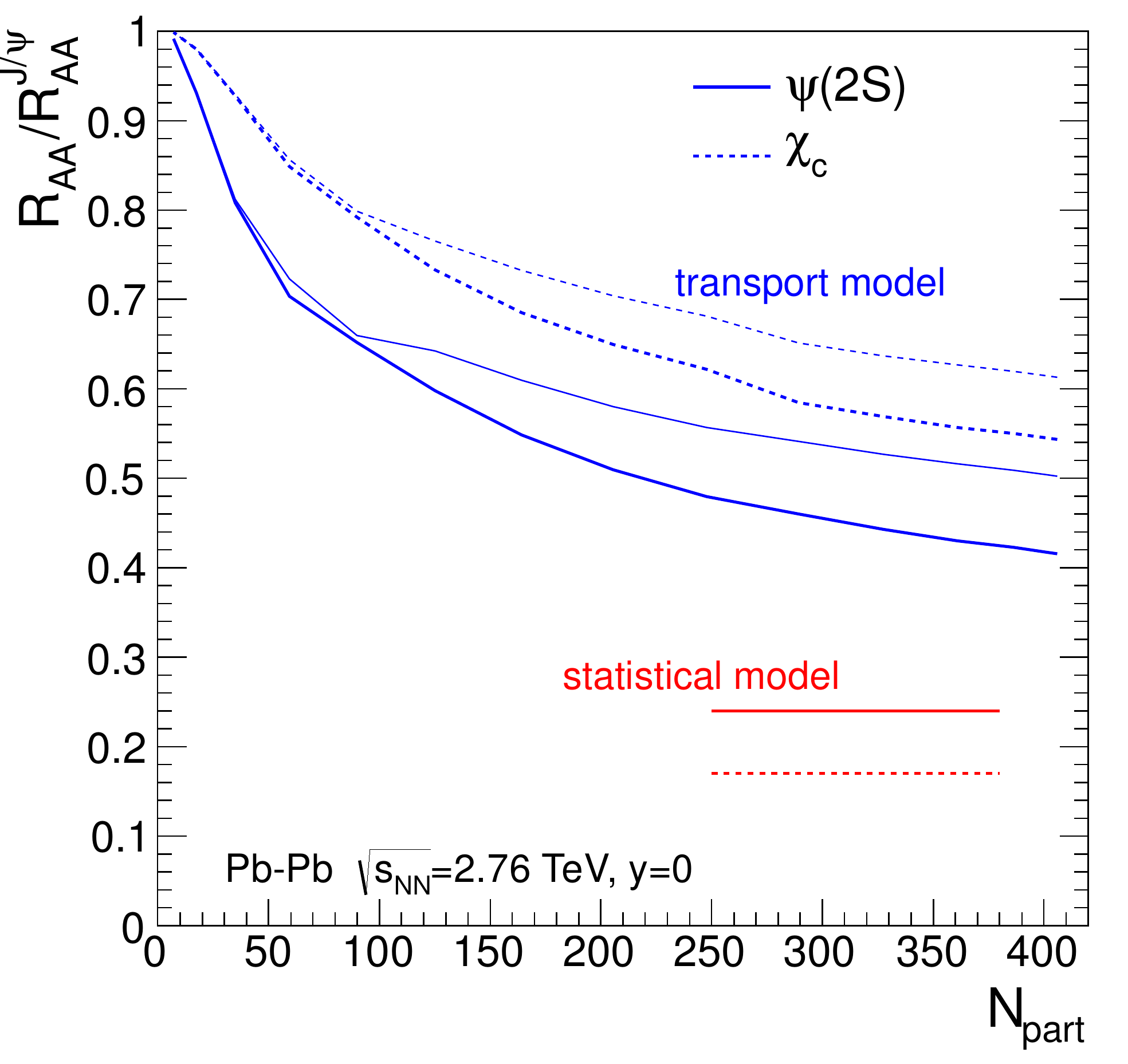}}
\end{minipage} \end{tabular}
\caption{Centrality dependence of the ratio of the nuclear modification factor 
of higher charmonium states to that of J/$\psi$ for the LHC energy.
The present data on $\psi(2S)$ are shown in the left panel (plot from 
\cite{CMS-PAS-HIN-12-007}). 
In the right panel are shown transport model \cite{Zhao:2011cv} and 
statistical hadronization model \cite{Andronic:2009sv}
predictions for $\psi(2S)$ (continuous lines) and $\chi_{c1,2}$ (dashed lines).
} 
\label{fig:raa_psi2s} 
\end{figure}

The production of the $\psi(2S)$ charmonium state was proposed early on 
as a signature of the QCD phase boundary \cite{Sorge:1997bg,BraunMunzinger:2000px} and also to possibly reveal in-medium effects on decay properties 
\cite{Grandchamp:2003uw}.
Present data on $\psi(2S)$ production at colliders \cite{Arnaldi:2012bg,CMS-PAS-HIN-12-007}, Fig.~\ref{fig:raa_psi2s} (left panel), is affected by rather 
large statistical uncertainties and also by imposed kinematic cuts to perform
the challenging measurement of $\psi(2S)$. This measurement offers the 
potential to discriminate between the statistical hadronization and the kinetic 
models, see Fig.~\ref{fig:raa_psi2s} (right panel).  
A significant progress on the experimental side is expected with the upgraded 
ALICE detector at the LHC \cite{Abelevetal:2014cna}.

The measurements of J/$\psi$ production in p--Pb collisions at the LHC 
\cite{Abelev:2013yxa,Aaij:2013zxa} exhibit features very different than
in Pb--Pb collisions \cite{Book:2014kta}.
The data are described by shadowing calculations 
\cite{Albacete:2013ei,Ferreiro:2013pua}, but not in all 
details, though. A saturation model \cite{Fujii:2013gxa} underpredicts the data.
The mechanism of energy loss in cold nuclear matter 
\cite{Arleo:2012rs}, in conjunction with shadowing effects, provides a better
description of the data. The implications of this
mechanism for Pb--Pb collisions has just been put forward \cite{Arleo:2014oha}.
Hot matter effects on J/$\psi$ production in p--Pb collisions have also been 
proposed \cite{Liu:2013via}.
The centrality (event activity) dependence of J/$\psi$ production exhibits
a host of features \cite{Lakomov:2014yga,Blanco:2014kta}, which are yet 
to be understood.
Interesting aspects are revealed by the measurement of $\psi(2S)$ production 
in d--Au collisions at RHIC \cite{Adare:2013ezl} and in p--Pb at the 
LHC \cite{Abelev:2014zpa,Arnaldi:2014kta}, indicating possible final-state 
effects.

\section{Bottomonium}
\label{bottom}

Recent measurements of the production of bottomonium ($b\bar{b}$) states at 
the LHC \cite{Chatrchyan:2012lxa,Chatrchyan:2013nza,Abelev:2014nua} and at 
RHIC \cite{Adamczyk:2013poh,Adare:2014hje} add an important new aspect to the 
quarkonium dissociation story.
The nuclear modification factor for the $\Upsilon$ states measured at the LHC 
at midrapidity was interpreted as evidence for the sequential suppression 
mechanism \cite{Chatrchyan:2012lxa}. 
In this picture, the yield of the $\Upsilon(2S)$ is expected to vanish, while
a non-zero yield is measured \cite{Chatrchyan:2012lxa}. 

\begin{figure}[hbtp]
\begin{tabular}{lr} \begin{minipage}{.49\textwidth}
\hspace{-0.3cm}\includegraphics[width=1.02\textwidth,height=.75\textwidth]{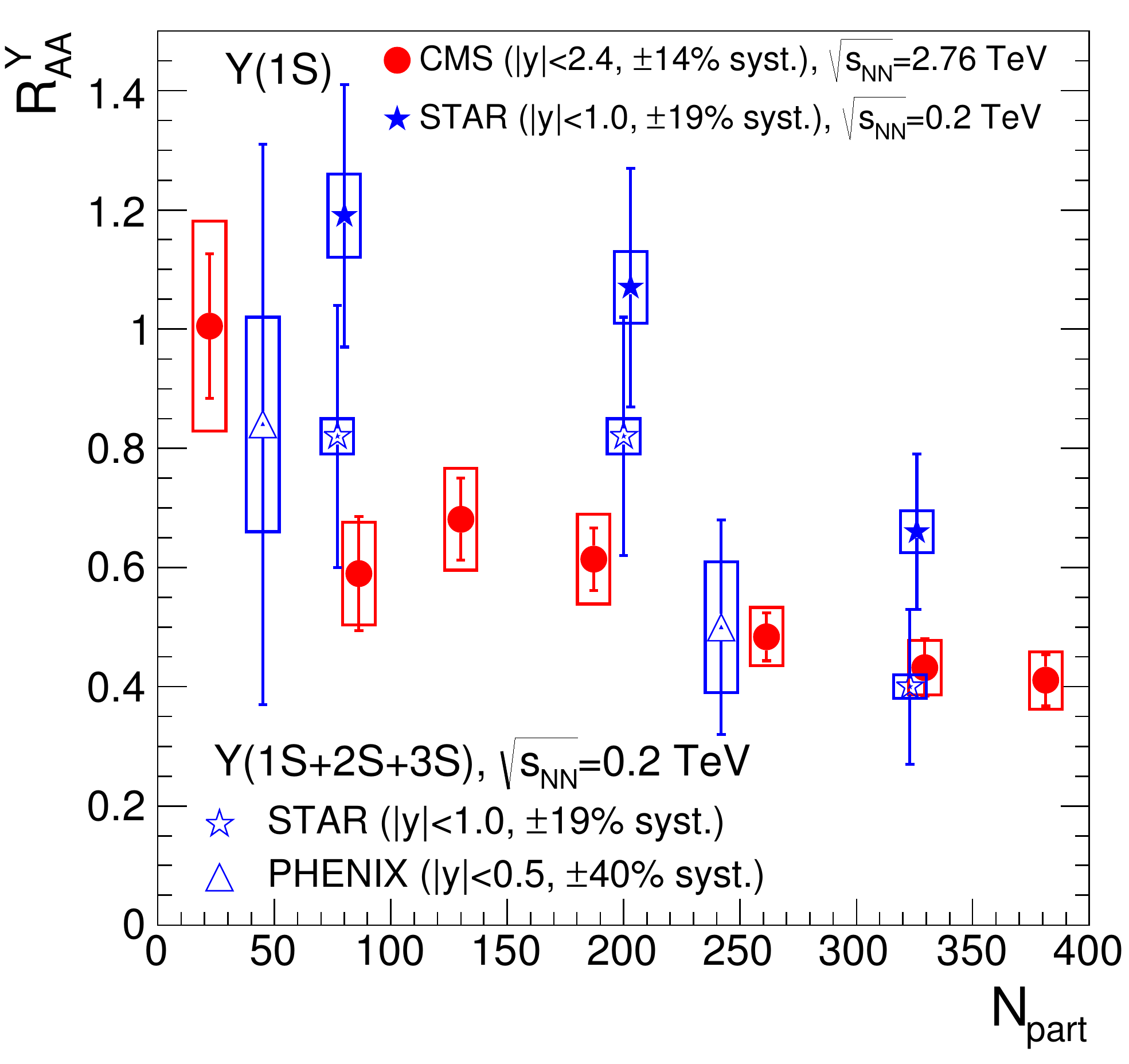}
\end{minipage} & \begin{minipage}{.49\textwidth}
\hspace{-0.3cm}\includegraphics[width=1.02\textwidth,height=.75\textwidth]{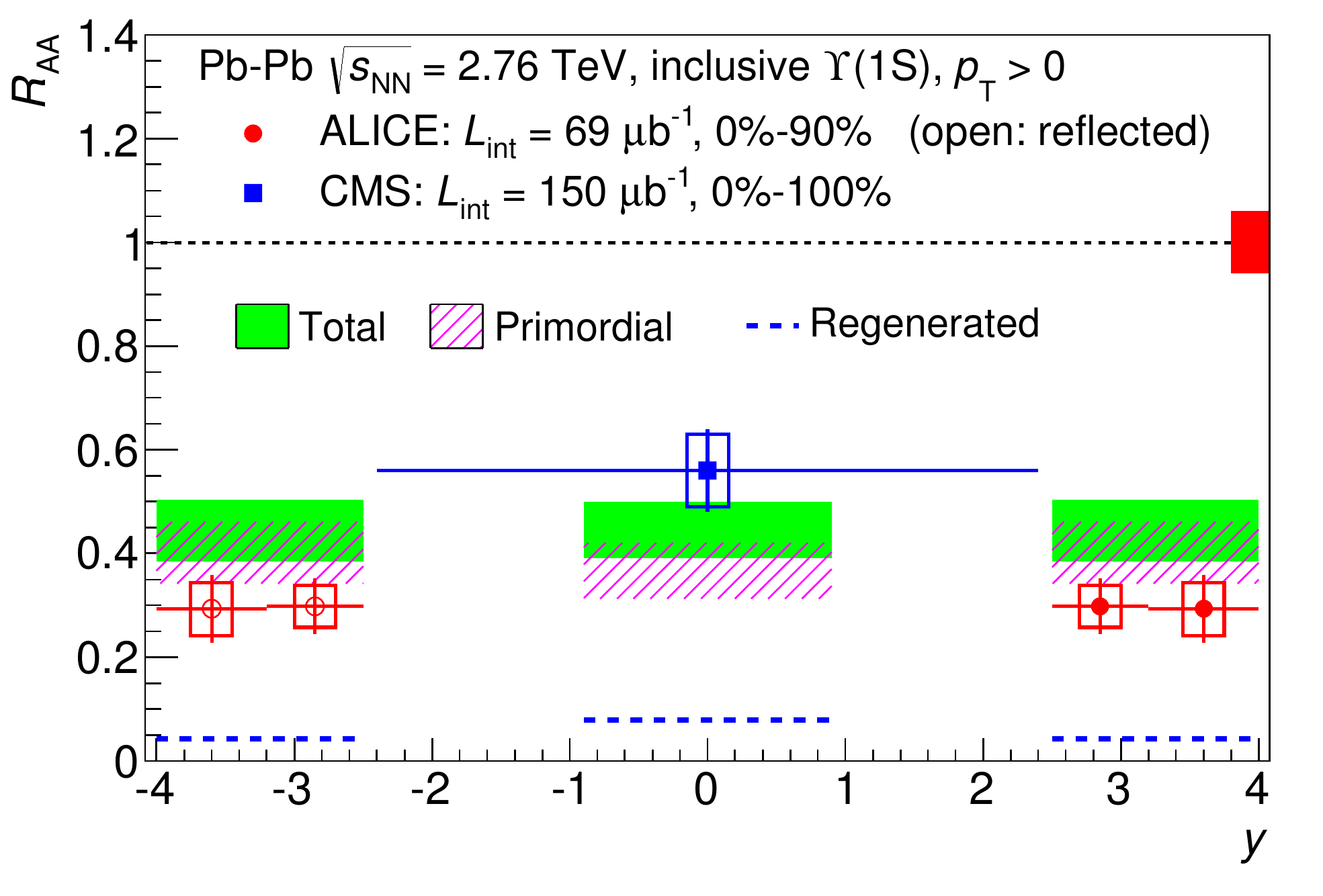}
\end{minipage}\end{tabular}
\caption{The dependence of the nuclear modification factor $R_{\mathrm{AA}}$ for
$\Upsilon$ production at the LHC on centrality (left panel) and on rapidity 
(right panel, plot from \cite{Abelev:2014nua}).
The data are integrated over $p_{\mathrm T}$  and are from the PHENIX 
\cite{Adare:2014hje} and STAR \cite{Adamczyk:2013poh} collaborations at RHIC 
and the CMS \cite{Chatrchyan:2012lxa} and ALICE \cite{Abelev:2014nua} 
collaborations at the LHC.
Note the additional systematic uncertainties of the data quoted in 
the legend. The transport model predictions in the right panel are 
from \cite{Emerick:2011xu}.
}
\label{fig:raa_Y}
\end{figure}

An $\Upsilon$ suppression similar in magnitude to that measured at the LHC 
is observed at RHIC \cite{Adamczyk:2013poh,Adare:2014hje}, see 
Fig.~\ref{fig:raa_Y} (left panel). 
This brings to mind an analogous pattern for J/$\psi$ 
at the SPS and at RHIC energies, already mentioned above: the magnitude of 
its suppression was observed to be similar up to the top RHIC energy.
The $\Upsilon$ suppression observed at RHIC is described 
by transport model predictions \cite{Liu:2010ej,Emerick:2011xu},
see model comparisons in \cite{Adamczyk:2013poh}.

The data on $\Upsilon(1S)$ production at forward rapidity \cite{Abelev:2014nua}
exhibit a stronger suppression than at mid-rapidity \cite{Chatrchyan:2012lxa}, 
see Fig.~\ref{fig:raa_Y} (right panel).
This recalls the observation at RHIC of a stronger suppression of
J/$\psi$ at forward rapidity than at midrapidity, which was interpreted as
evidence for J/$\psi$ production via the statistical hadronization 
mechanism \cite{Andronic:2007bi}.

The $\Upsilon(1S)$ data at the LHC indicate that the Debye screening mechanism, 
implemented in a hydrodynamical approach \cite{Strickland:2012cq}, does not 
describe the measurements \cite{Abelev:2014nua}.
Transport model predictions \cite{Emerick:2011xu} describe the features of
the LHC data \cite{Abelev:2014nua}, see Fig.~\ref{fig:raa_Y} (right panel), 
exhibiting though a much weaker rapidity dependence than the data.
They indicate a rather small (re)generation component for $\Upsilon(1S)$ and a 
large primordial production \cite{Emerick:2011xu,Zhou:2014hwa}; 
$\Upsilon(2S)$ production arises in transport models exclusively from 
(re)generation \cite{Zhou:2014hwa}. 

\begin{figure}[htbp]
\centerline{\includegraphics[width=.55\textwidth]{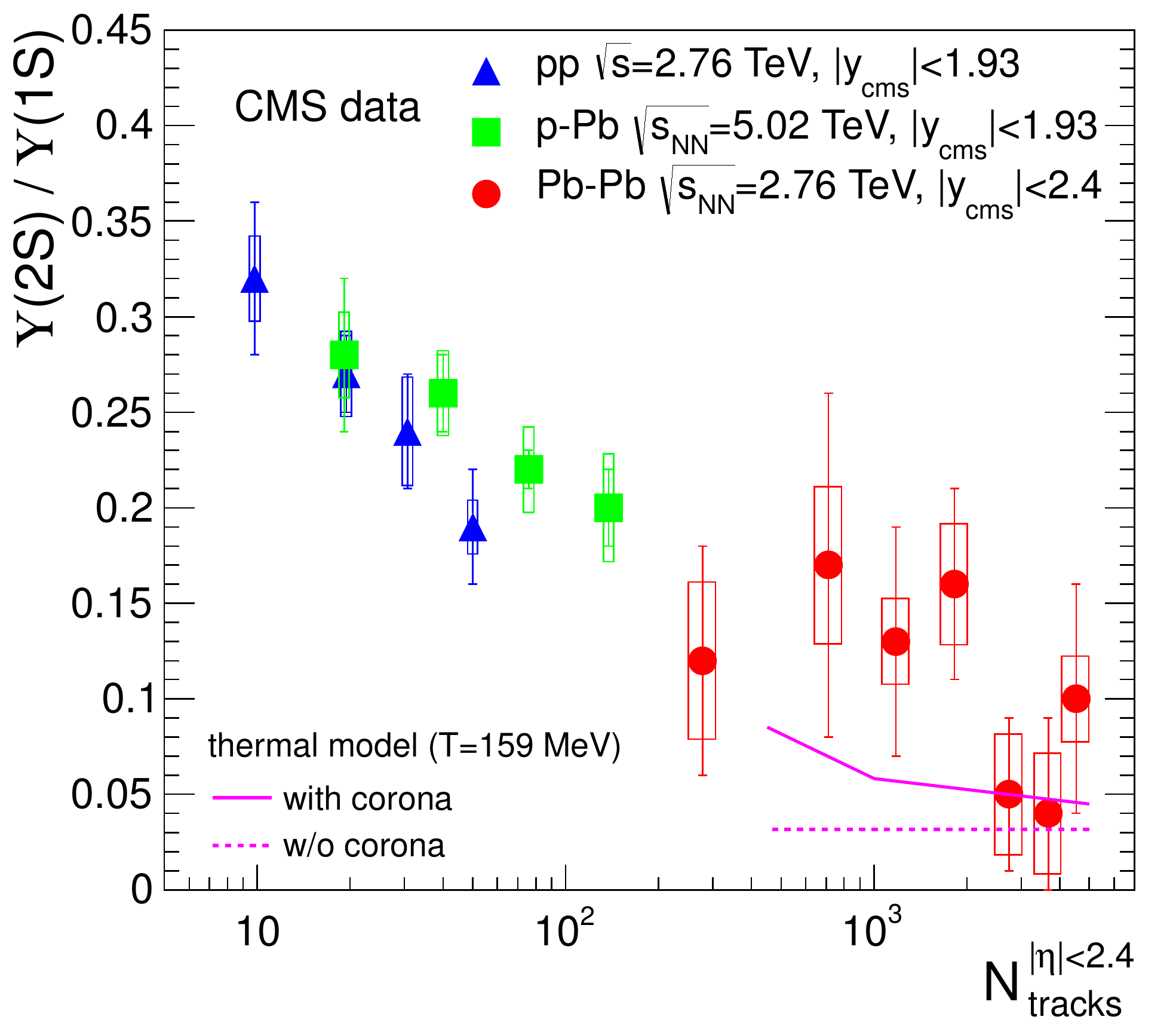}}
\caption{Track multiplicity dependence of the ratio of yields of
 $\Upsilon(2S)$
and $\Upsilon(1S)$ bottomonium states, measured at the LHC in pp, p--Pb and 
Pb--Pb collisions by the CMS collaboration \cite{Chatrchyan:2013nza}. 
The lines are thermal model predictions for central Pb--Pb collisions; 
the full line includes an estimate of the contribution of the production in the 
corona of the colliding nuclei.
} 
\label{fig:y} 
\end{figure}

The production ratio $\Upsilon(2S)/\Upsilon(1S)$ is expected in the 
statistical model \cite{Andronic:2009sv} to be very different in Pb--Pb 
compared to pp collisions.
Such an expectation is exhibited by the recent data at the LHC 
\cite{Chatrchyan:2013nza}, shown in Fig.~\ref{fig:y}. A gradual decrease
of the yield ratio $\Upsilon(2S)/\Upsilon(1S)$ is seen as a function
of track multiplicity in pp, p--Pb and Pb--Pb collisions, a trend which
is yet to be understood.
The data are compatible, for central Pb--Pb collisions, to the value predicted 
by the statistical hadronization model for $T=159$ MeV.
This provides a tantalizing possibility of adding the bottom flavor towards
constraining even further the QCD phase boundary with nucleus-nucleus data at 
high energies.

\section{Conclusions and outlook}

The charmonium data in Pb--Pb collisions at the LHC, in conjunction with data
at RHIC, show that (re)generation in deconfined matter during the QGP lifetime 
or generation at the chemical freeze-out (hadronization) appear to be the only
mechanisms of production that describe the measurements.
While rather convincing at the LHC, the mechanism of generation by statistical 
hadronization may be disputed for RHIC (and SPS) energies, where transport 
models currently indicate a fraction of (re)generated charmonium below 50\%.
Even as it remains important to clarify all details, it seems safe to affirm,
based on models, that we do see in experiments charmonium suppression in 
deconfined matter.
The LHC data on bottomonium suppression were followed very recently by data 
at RHIC, where the measurement is clearly much more challenging. 
More theoretical insight is awaited in the bottomonium sector.

Even as ``generation" and ``(re)generation" seem equivalent terms,
they label two rather different pictures of quarkonium production; 
discriminating between them will help providing answers to fundamental 
questions related to the fate of quarkonium in a hot, deconfined, 
medium \cite{Liu:2006nn,Mocsy:2007yj,Laine:2011xr}. 
Is any hadron having a chance of survival in the hot and dense soup of 
deconfined quarks and gluons? 
Do charm (or bottom) quarks hadronize at the same temperature as their 
lighter siblings \cite{Bellwied:2013cta,Bazavov:2014yba}?
Quarkonium data at RHIC and in particular at the full LHC energy
will hopefully help clarifying such questions in the next years, providing a
better understanding of deconfined QCD matter.
Obviously, sustained effort on the theoretical front is needed along the way.
New theoretical ideas continue to appear \cite{Kharzeev:2014pha}.

\section*{Acknowledgments}

I would like to thank the RHIC and LHC collaborations for making available 
their results in a very accessible way. 
For discussions and clarifications about the data I thank R. Arnaldi, 
I. Arsene, C. Blume, G. Bruno, J. Castillo, T. Dahms, R. Granier de Cassagnac, 
B. Hong, D. Morrison, J. Nagle, E. Scomparin, C. Suire, M. Winn, and N. Xu.
For enlightening discussions on theoretical models I am grateful to 
P. Braun-Munzinger, R. Rapp, K. Redlich, J. Stachel, X. Zhao, and P. Zhuang.


\bibliography{quonium} 

\end{document}